\documentclass[fleqn,10pt]{wlscirep}
\usepackage{graphicx}
\usepackage{dcolumn}
\usepackage{bm}
\usepackage{subfigure}
\usepackage{epsfig}
\usepackage{color}

\begin{document}
\title{Clusters in sedimentation equilibrium for an experimental hard-sphere-plus-dipolar Brownian colloidal system} 
\author[1]{Hugh D. Newman}
\author[1,*]{Anand Yethiraj}
\affil[1]{Department of Physics and Physical Oceanography, Memorial University, St. 
John's, NL, A1B 3X7, Canada}
\affil[*] {ayethiraj@mun.ca}
\begin{abstract}
In this work, we use structure and dynamics in sedimentation equilibrium, in the presence of gravity, to examine, $via$ confocal microscopy, a Brownian colloidal system in the presence of an external electric field. 
The zero field equation of state (EOS) is hard sphere without any re-scaling of particle size, and the hydrodynamic corrections to the long-time self-diffusion coefficient are quantitatively consistent with the expected value for hard spheres.
Care is taken to ensure that both the dimensionless gravitational energy, which is equivalent to a Peclet number $Pe_g$, and dipolar strength $\Lambda$ are of order unity.
In the presence of an external electric field, anisotropic chain-chain clusters form; this cluster formation manifests itself with the appearance of a plateau in the diffusion coefficient when the dimensionless dipolar strength $\Lambda \sim 1$. The structure and dynamics of this chain-chain cluster state is examined for a monodisperse system for two particle sizes.
 
\end{abstract}
\maketitle
\section*{Introduction}
Electric-field-induced dipolar colloids are a simple two-component system where both cluster formation and network formation \cite{tao_electric_1989, klingenberg_structure_1989, martin_structure_1998, barhoum_new_2013, kumar_new_2005, agarwal_low-density_2009} have been observed during phase separation. 
Using external electric fields, colloid phase behavior can be modified to enable field-switchable phase transitions from fluid and close-packed-crystalline, to structures that are anisotropic along the field direction. The resulting electrorheological effect has been well studied experimentally \cite{winslow_induced_1949, tao_electric_1991, martin_structure_1998}, and can form a basis for studying the kinetics of crystal transformations \cite{mohanty_multiple_2015}. Systems of monodisperse colloids with an effective dipolar interaction induced by an external electric field have been demonstrated for charge-stabilized (hard-sphere-like) silica colloids in aqueous suspension, for colloids with strong electrostatic repulsions (PMMA in non-aqueous media) and for ultrasoft microgel colloids in aqueous suspensions as well \cite{yethiraj_monodisperse_2002, mohanty_deformable_2012}.

Computer simulations of colloidal hard spheres with an imposed pair-wise electric polarization interaction that neglect contributions of higher order than the point-dipolar term seem to capture the essence of the behaviours \cite{klingenberg_structure_1989, hynninen_phase_2005}. While steady-state clusters are seen in both experiments \cite{dassanayake_structure_2000, agarwal_low-density_2009} and computer simulations \cite{almudallal_simulation_2011, hynninen_phase_2005}, an unusual ultra-low density network-forming phase at very low densities ($\Phi < 0.04$) (referred to as the ``void phase'') has been observed in independent experiments \cite{kumar_new_2005, agarwal_low-density_2009} but not observed in simulations \cite{almudallal_simulation_2011, richardi_low_2011, park_electric-field-induced_2011}. This has raised the question whether the large-scale structures are equilibrium or out-of-equilibrium structures.  Another possibility is that the discrepancy lies in the quantitative details of the inter-particle potential. A third factor is the role of gravity in experiments, which is unaccounted for in simulations. Simulations of dipolar hard spheres have focused on regimes (high field and/or high packing fractions) where crystals form, because that is primarily where past experiments have focussed. Less work has been in the regimes of simultaneously low packing fraction and low dipolar strengths (of order $k_B T$). 

%
\begin{figure}[th]
\centering
\includegraphics[width=85mm]{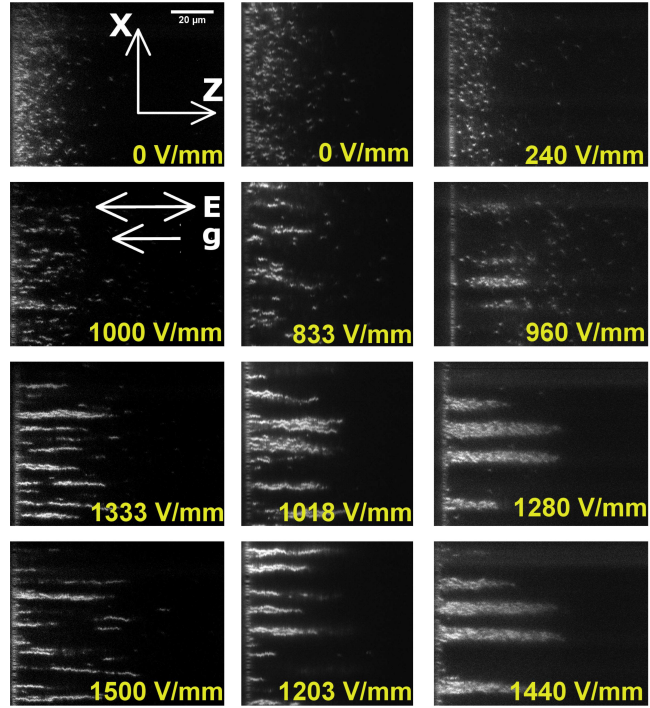}
\caption{Effect of an electric field in the presence of gravity. Left panel: 0.8$\mu m$ diameter PMMA colloids in 70:30 decalin/TCE. Snapshots ($x$-$z$) at four selected fields are shown (gravity to the left, and  $z$ increasing to the right). Above a critical field, $E > 1000$V/mm, there are well-formed chains. The left hand corner of each image is the substrate, and $z$ increases to the right. Middle panel: 1.0$\mu m$ diameter PMMA colloids in 60:40 decalin/TCE. Snapshots ($x$-$z$) at 4 selected fields (with comparable dipolar strengths $\Lambda$). Right panel: The binary mixture has the thickest columns of chains.}
\label{fig:exp13_xz}
\end{figure}
{
One can minimize the effects of gravity by density matching, by ensuring that the gravitational length, defined by  $l_g = k_BT/(4\Delta \rho g a^3/3)$ is much greater than the particle diameter, i.e. the ratio of gravitational potential energy to thermal energy, given by $Pe_g = 2a/l_g << 0.1$. Since $Pe_g$ scales as $a^4$, particles twice this size, typical in many microscopy experiments, would have $Pe_g\sim O(1)$, and so even with careful density matching gravity is hard to neglect. 
}

{Other forms of aggregation have been studied extensively in colloidal systems \cite{poon_phase_1998}, particularly in the presence of polymer-induced depletion attractions. Phase separation, gelation and glassy behaviour have been observed in systems with a combination of short-range attractions in addition to (hard-sphere or electrostatic) repulsions (see the review by Zaccarelli \cite{zaccarelli_colloidal_2007}), resulting in different rheological properties \cite{koumakis_two_2011}. Stable cluster states have been identified in {\it equilibrium} colloidal systems with competing interactions on different lengthscales \cite{stradner_equilibrium_2004, barhoum_new_2013}. Understanding these cluster states remains a problem of fundamental interest.
}

{In this work, we employ a colloidal system to study dipolar clusters. Our system is easily index matched and density matched for confocal microscopy experiments, and we show it to be an excellent candidate for true hard-sphere-plus-dipolar interactions. With colloidal spheres of diameter $2a = 0.8$ and 1.0 $\mu$m-diameter in a nearly density-matched solvent mixture, our system, with $Pe_g \sim 0.1$ is as strongly Brownian as possible while still allowing the particles to be just large enough for microscopy. In our experiments, there is a 20\% difference in the two particle sizes used, which is significant given the $a^4$ dependence of the Peclet number, and much larger than the (less than) 5\% polydispersity in particle size.
}

Sedimentation in the presence of gravity has been used to good effect in many colloidal systems to extract information about equations of state; an early example is from Hachisu and Takano \cite{hachisu_pressure_1982} while a review by Piazza \cite{piazza_settled_2014} summarizes recent developments. In this study, we first demonstrate that our Brownian colloidal system has interactions that are hard-sphere-plus-dipolar. Then we examine structures and dynamics of clusters in sedimentation equilibrium, and characterize the dependence of cluster size and dynamics as a function of height, for two particle sizes.

\section*{Results}

Experiments were carried out for two monodisperse colloidal systems with 0.8 and 1 $\mu$m-diameter spheres, as well as a bidisperse system, at several values of a uniaxial AC electric field (see Methods).
Snapshots of colloidal sediments at zero field and some selected field strengths are shown in Figure~\ref{fig:exp13_xz}. The alternating electric field is along $z$, i.e. parallel and anti-parallel to the direction of gravity (which points to the left). Figure~\ref{fig:exp13_xz} (left panel) shows 4 x-z particle profiles for the 0.8$\mu m$ diameter PMMA colloids in 70:30 decalin/TCE. As the field increases from zero to 1667 V/mm, the sediment goes from fluid-like to fully-formed chains. The middle panel in Figure~\ref{fig:exp13_xz} shows comparable profiles for 1.0$\mu m$ diameter PMMA colloids in 60:40 decalin/TCE. The field strengths chosen are ones for comparable values of the dipolar strength 
\begin{equation}
\Lambda = \pi \epsilon_0 \epsilon_f \beta^2 a^3 E_0^2/(2 k_B T),
\label{eqn:dipstr}
\end{equation}
 where $\beta = (\epsilon_p - \epsilon_f)/(\epsilon_p +  2\epsilon_f)$, $\epsilon_p$ and $\epsilon_f$ are the particle and fluid dielectric constant respectively, $a$ is the particle radius, and $E_0$ is the amplitude of the sinusoidal AC electric field; see Table~\ref{tab:lambda} in Methods for a numerical relation between $E_0$ and $\Lambda$). For comparison, Figure~\ref{fig:exp13_xz} (right panel) shows profiles for a bidisperse system composed of 0.8$\mu m$ and 1.0$\mu m$ diameter particles (only the smaller  0.8$\mu m$ particles are visible in this image) in 50:50 ratio by volume fraction. The particle columns in the bidisperse system are always thicker: this is even clearly the case when only half the particles (by volume fraction) are visible.

%
\begin{figure}
     \centering
     \includegraphics[width=160mm]{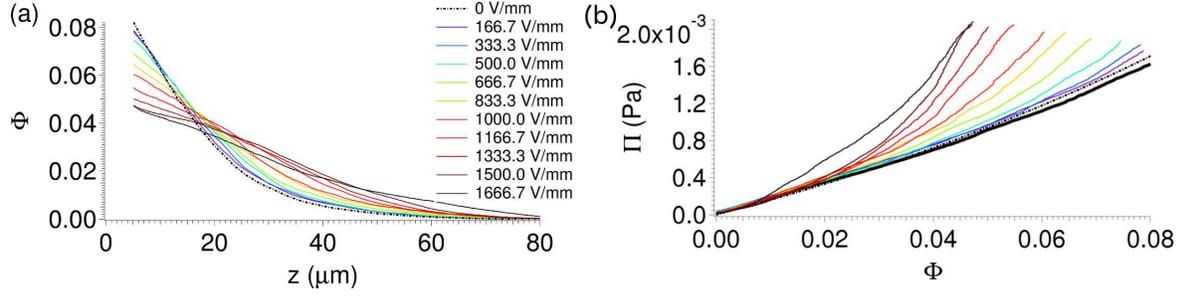}
     \caption{Effect of an electric field ($E=0$ V/mm to $E=1666.7$V/mm) on sedimentation equilibrium. 0.8$\mu$m diameter PMMA colloids in 70:30 decalin/TCE in the presence of gravity. (a)
Sedimentation profiles become progressively more extended as a function of the field strength.
(b) The zero-field equation of state (field strengths are as in the legend in (a))
is consistent with hard spheres (the solid black line is the Carnahan-Starling equation of state), while the osmotic pressure is increasingly higher than the hard-sphere value with increasing field strength.
}
  \label{fig:exp13_phivsz_all}
\end{figure}

Figure~\ref{fig:exp13_phivsz_all}(a) shows the sedimentation profiles, $\Phi$ as a function of $z$, for all fields for the 0.8$\mu$m-diameter colloids (mean volume fraction $\Phi = 0.017$), obtained from the 3-dimensional image stacks.
We see in Figure~\ref{fig:exp13_phivsz_all}(a) that the profiles are more extended as the field increases. This is expected due to the onset of the formation of string-like clusters of particles, which we already saw in Figure~\ref{fig:exp13_xz}. 
One can also calculate the equation of state by simultaneously obtaining the local volume fraction $\Phi(z)$ in thin slices $dz$ at height $z$, parallel to and above the bottom substrate and the local osmotic pressure $\Pi(z) = \int_z^\infty g \Delta \rho \Phi(z') dz'$ as a function of field strength is shown in Figure~\ref{fig:exp13_phivsz_all}(b) \cite{ning}. 
We can see that the zero field equation of state (EOS) for $0.8 \mu m$ PMMA particles shows excellent agreement with what is expected for a hard-sphere EOS from the Carnahan-Starling relation, and is in agreement with careful experiments in true bulk (non-microscopy) systems~\cite{piazza_equilibrium_1993, rutgers_measurement_1996}. This shows that our system is an excellent hard-sphere system, at least in the dilute regime which is of interest here. While the packing fraction range for these experiments is low, it has been seen in previous work that deviation from hard-sphere-like behaviour, for example due to a soft particle shell, manifests itself in a larger effective particle diameter {\it even at low concentrations}. For example, in the work of Li {\it et al} \cite{ning}, which was a silica in water-DMSO colloidal system where the void phase was observed \cite{agarwal_low-density_2009}, the EOS could only be fit by assuming an effective particle diameter that was 20\% larger than the measured value. 

{We can also obtain the osmotic pressure from the sedimentation profiles at non-zero fields; this is shown in Figure~\ref{fig:exp13_phivsz_all}(b). We note that the osmotic pressure obtained is only universally valid for single particles and small chains because the effective Peclet number increases quickly with the size of the structure; we thus refer to it as an {\it apparent} osmotic pressure $\Pi$. It is nevertheless a useful quantity because the intra-chain structure is the result of both gravitational force and free thermal exchange of particles when $\Lambda$ is O(1). 
}
At non-zero fields (i.e. with increasing $\Lambda$), the key observation is that $\Pi$ at any given packing fraction $\Phi$ is larger than the hard-sphere value, and increases with the field. This is in contrast with experiments (Li {\it et al} \cite{ning}) where the dipoles were perpendicular to gravity, where $\Pi$ at a given $\Phi$ decreased with increasing $\Lambda$. Pressure in a fluid is isotropic. The {\it apparently} anisotropic osmotic pressure perhaps arises from our definition of the local volume fraction. While the mean volume fraction decreases continuously with $z$, as shown in Figure~\ref{fig:exp13_phivsz_all}(a), the particles in any given slice along the $x$-$y$ plane are not homogenously distributed: there is clustering, so the true local volume fraction is higher than the mean value. Thus we next focus on the properties of the clusters.

\begin{figure}
     \begin{centering}
     \includegraphics[width=160mm]{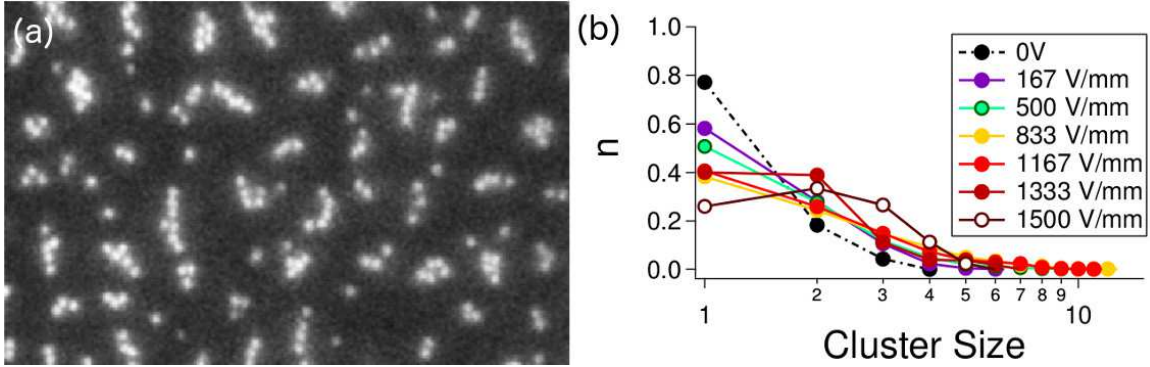}
     \end{centering}
     \caption{(a) Lateral ($x$-$y$) snapshot of chain-chain clusters for 0.8$\mu m$ diameter PMMA colloids in 70:30 decalin/TCE at $z=40\mu m$, at a field E = 1500 V/mm.
(b) Distribution of lateral size (in units number of chains) corresponding to the slice above at $z=40\mu m$. Electric field and gravity point into the page (i.e. along $z$). Clusters of size 2 and larger become more frequent for the higher fields.
}
  \label{fig:cluster}
\end{figure}

Figure~\ref{fig:cluster}(a) shows clusters a height of $40\mu m$ above the bottom substrate. There is a distribution: there are several individual chains, doublets, triplets, and elongated multi-chain clusters. At zero field (the black points in Figure~\ref{fig:cluster}(b)), there are single-particles and some two-particle clusters. These are not real clusters. Since a cluster is defined simply by proximity ($r \le 1.1 \times 2a$), the probability of unassociated two-particle clusters is not zero.
On the other hand, as the field is increased, the probability of size 2, 3, and larger clusters increases systematically above the zero-field value; this represents real in-plane clustering (along $xy$) of chains (along $z$).

\begin{figure}
     \centering 
     	\includegraphics[width=160mm]{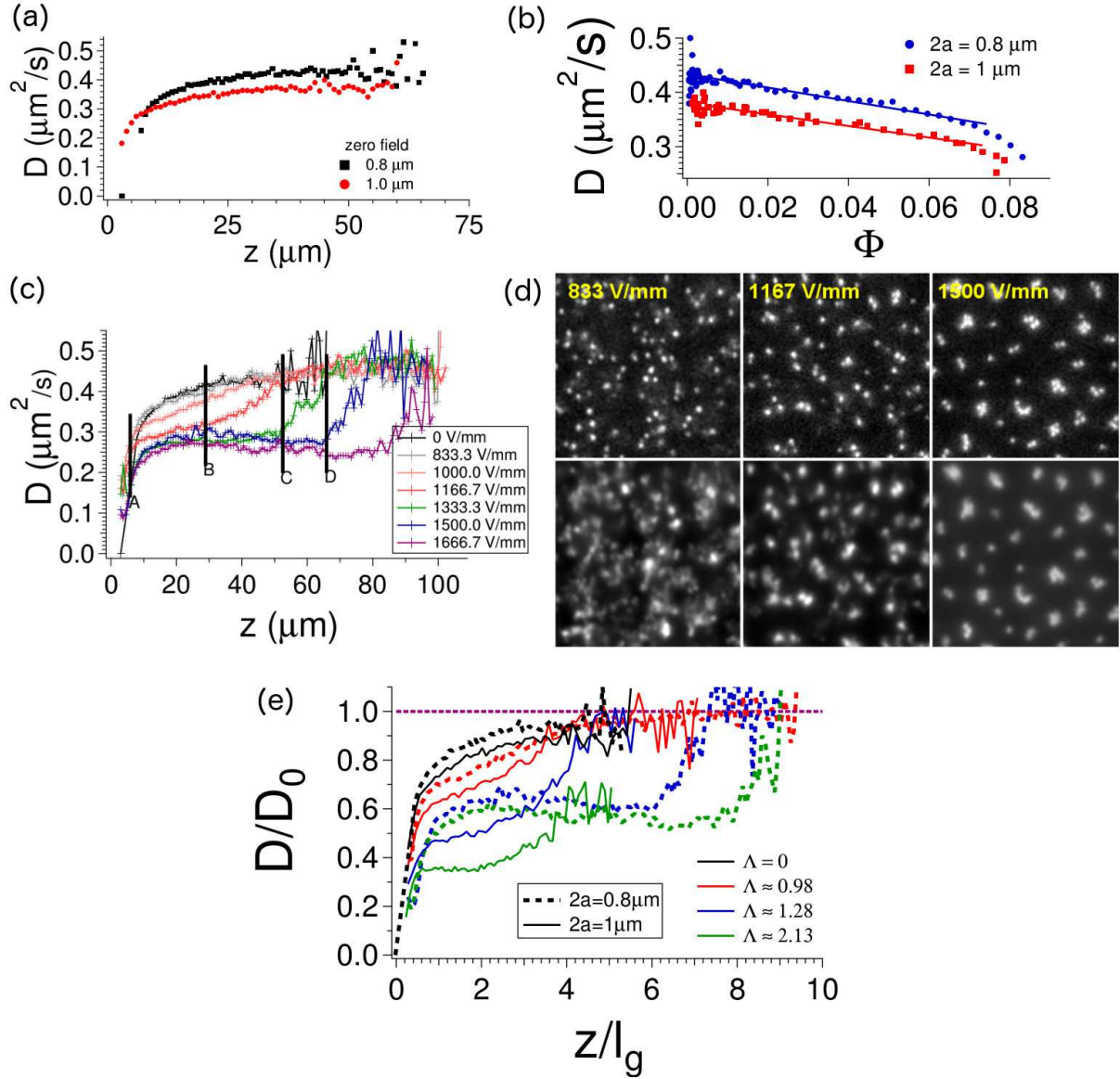}
     \caption{Diffusion as a function of $z$ for a range of field strengths from $E=0 V/mm$ to $E=1666.7 V/mm$ of 0.8$\mu m$ diameter PMMA colloids in 70:30 decalin/TCE. At zero field, diffusion coefficients (for both monodisperse systems ($2a=1\mu m$ PMMA in 60:40 decalin/TCE and $2a=0.8\mu m$ PMMA in 70:30 decalin/TCE) (a) increase towards the bulk value with increasing height, and (b) decrease (linearly) with local volume fraction $\Phi$, yielding both the diffusion coefficient at infinite dilution $D_0$ and the hydrodynamic factor $K^L$.
(c) The diffusion coefficient varies with height $z$ due to effects of varying concentration as well as wall effects.
     (d) Snapshots (top panel) from movies taken at three different fields at $z = 28 \mu$m, corresponding to position B in (c) show the onset of clustering. The time-averaged intensities (bottom panel) show that particles are much more dynamic at low fields, while they are confined to chain clusters at high fields (see text).
(e) A plateau emerges at $\Lambda \sim 1$ in the value of scaled diffusion coefficients for both monodisperse systems plotted against the dimensionless $z/l_{g}$.}
  \label{fig:diffall}
\end{figure}

Evidence of clustering can also be obtained from ``sedimentation-diffusion''  profiles where we measure the lateral (in-plane) self-diffusion coefficient $D$ (over a few seconds, which is already in the long-time limit for single-particle motion) as a function of distance $z$ from the substrate. 
At zero field, the diffusion coefficient $D$ saturates to the bulk value $D_0 \sim 0.44 {\mu \mathrm{m}}^2/s$ when $z$ is greater than 50 ${\mu \mathrm{m}}$. We can utilize the diffusion profiles in Figure~\ref{fig:diffall}(a) and the sedimentation profiles in Figure~\ref{fig:exp13_phivsz_all}(a) to obtain the diffusion coefficient as a function of the volume fraction $\Phi$: this is shown in figure~\ref{fig:diffall}(b). For both particle sizes, the relation has the form $D = D_0(1 + K^L \Phi)$ with $K^L = -2.80 \pm 0.05$. This is comparable to the range -2.8 to -2.9 obtained in experiments by Blees {\it et al} \cite{blees_self-diffusion_1996} and de Kruif {\it et al.} \cite{kruif_physics_1987} but different from the theoretical value of -2.1 obtained by Batchelor \cite{batchelor_brownian_1976, russel_colloidal_2001}. This shows that the density gradient imposed for our experiments does not affect the hydrodynamics, because we get essentially the bulk experimental value for long-time diffusion.

We next examine the effect of a field in different regions of the sediment. At position A in Figure~\ref{fig:diffall}(c), substrate effects are strong, and increasing the field slows the dynamics significantly due to the formation of chain clusters - this can already be seen at the lowest fields, e.g. in the second row of Figure~\ref{fig:exp13_xz}. At position B  ($z =30 \mu m$ in Figure~\ref{fig:diffall}(c)), the diffusion coefficient decreases from $D = 0.42 {\mu \mathrm{m}}^2/s$ to $0.27 {\mu \mathrm{m}}^2/s$ as the field increases from zero to E = 1667 V/mm. At high fields, there is a plateau for $D$ versus $z$, which is an indicator that all particles are associated with chain-like clusters. 
At intermediate fields (E = 1000 and 1167 V/mm) the dynamics is intermediate between the high-field plateau and the zero field value, which signals the end of a chain-like cluster and the beginning of the diffuse particle region seen in Figure~\ref{fig:exp13_xz}.
This is also seen in Figure~\ref{fig:diffall}(d) (top panel), corresponding to $z = 28 \mu$m (position B) where sharp clusters are seen at the highest fields, and in time averages of the movies (bottom panel), where the particle dynamics smear the intensities at lowest fields, but the clusters are still visible at highest fields, suggesting particles stay confined to their respective clusters. At intermediate fields, the time average (bottom panel, middle) is more diffuse than the snapshot (top panel, middle), but the clusters are still visible, indicating that the dynamics is also intermediate between free ``particle'' and cluster ``particle'' (note that the particles referred to here are chains along $z$).
At E = 1333 V/mm, the plateau extends to position C, while at 1500 V/mm, it extends further to position D, and even further at higher fields; i.e. the width of the plateau in the sedimentation-diffusion ($D$ versus $z$) profile increases with field, which is consistent with the increase in chain lengths with increasing field, seen in Figure~~\ref{fig:exp13_xz}.

We compare these behaviours for the two particle sizes in Figure \ref{fig:diffall}(e). While the 0.8 $\mu$m particles exhibit a plateau at approximately $D/D_0 \sim 0.6$ and at $\Lambda \sim$ 1, the 1 $\mu$m particles only show the emergence of a plateau at $\Lambda \sim$ 2. At these larger dipolar strengths, the dynamics is slower, and $D/D_0 \sim 0.4$. The $D$ versus $z$ plateau is interesting. A possible picture for its existence is that particles that are associated with a chain are diffusing in the presence of a local potential well binding it to the chain cluster. Another interesting aspect, seen in Figure~\ref{fig:diffall}(d) (bottom panel) is that increasing the field effectively increases the strength of the lateral confinement. The current experiments have not probed large-scale motions on very long times, which requires also tracking the colloids in 3 dimensions. A systematic study of particle dynamics as a function of this confinement would be an interesting subject for future work.

Finally, we can assess the magnitude of the dipolar strength at the transition from single particles to clusters. 
{For the chain-cluster (or ``string fluid'') structure, the hard-sphere-plus-dipolar model \cite{hynninen_phase_2005} predicts that at packing fractions less $\Phi = 0.5$, the string fluid exists without showing clear body-centred tetragonal (BCT) crystallinity. All the structures seen at the low packing fractions of our current experiments show no crystallinity. In the simulations, chain clusters form at $\Lambda \le 4$ (note that $\gamma$ in the simulations \cite{hynninen_phase_2005} corresponds to 2$\Lambda$), in the experiments at $\Lambda \sim 1$. Experiment and the hard-sphere-plus-dipolar model are thus (at least qualitatively) in agreement. The results reported here now enable more detailed statistical comparisons to the hard-sphere-plus-dipolar model.} 
In addition, Figure \ref{fig:diffall} shows the scaled diffusion coefficient for both particle sizes, plotted against the height (scaled with the gravitational length). It is seen that the transition from single particles to clusters (i.e. the emergence of the lower plateau at approximately $D/D_0 \sim 0.6$) occurs at a dipolar strength $\Lambda$ that is close to 1 for both particle sizes (dotted and solid blue curves, referring to 0.8 and 1 $\mu$m spheres respectively). This again suggests that a pure-dielectric dipolar interaction is a dominant contributor to the interactions.

\section*{Discussion}
%
In this work, we have examined Brownian colloidal systems where the equation of state is shown to be in quantitative agreement with the hard-sphere equation of state, even in a thin sediment, and with no use of ``effective'' particle sizes. In addition, the hydrodynamic corrections to the diffusion coefficient at finite volume fractions are well captured by the Batchelor form for hard spheres. Thus, $\mu$m-scale colloids in a solvent mixture of decahydronapthalene and tetrachloroethylene are hard-sphere-like, density- and refractive-index-matchable, and yet have a strong enough dipolar inter-particle interaction in the presence of an external field. We used sedimentation-diffusion profiles, in zero field, to obtain self-diffusion coefficients as a function of volume fraction $\Phi$. The colloidal systems exhibited a decrease in self-diffusion coefficient with increasing $\Phi$ that was consistent with bulk experiments for hard-sphere colloids.
 
Next, we applied external electric fields. The ``hard-sphere-plus-dipolar'' equation of state (EOS) is obtained: the {\it apparent} osmotic pressure at any given volume fraction $\Phi$ is found to be greater than the hard-sphere value. It should be noted that, in the presence of clustering, the true local volume fraction is larger than the mean value at a particular height, so the EOS must be taken with a pinch of salt; however, we reiterate the observation in the raw sedimentation profiles that the profiles are more extended with higher fields.

We characterized the onset of field-induced interactions by sedimentation-diffusion profiles. The deviation from zero field profiles occurs at $\Lambda \sim 1$ (see Table~\ref{tab:lambda}). Stable anisotropic clusters form which are long along the field direction, but have a modest lateral extent. The peak in lateral size of clusters shifts to larger values with increasing field strength. At low fields, the self-diffusion coefficient increases steadily with height above the substrate $z$. At a dipolar strength of order unity, an interesting plateau develops develops in the $D$ vs $z$ profile. Snapshots of the clusters in this plateau region show well defined chain-like clusters indicating that the dynamics here corresponds to the internal dynamics of a (wagging or breathing) chain cluster. The length of the plateau correlates with the length of the chain clusters.

The lateral extent of chain-chain clusters increases with the dipolar strength. This is unsurprising, but the nature of the clusters is interesting: the clusters are thicker for $1 \mu m$-diameter particles than for $0.8\mu m$-diameter particles, as can be seen even by a visual inspection of Figure~\ref{fig:exp13_xz}. Also apparent here 
is that clusters are thickest in bidisperse suspensions, suggesting a synergistic effect of bidispersity. This apparent synergy likely arises because particle bidispersity works to the break the registry between adjacent chains. 

For context, two close-to-Brownian colloidal systems used in optical microscopy in which dipolar colloidal interactions have been previously employed are either at Peclet numbers of order 10, or not true hard-sphere systems. In the first system, silica microspheres in aqueous suspensions \cite{dassanayake_structure_2000, yethiraj_monodisperse_2002, agarwal_low-density_2009}, the particles are typically twice as dense as the medium, resulting in Peclet numbers (for micron-sized colloids) of order 10. In the second system, PMMA colloids in a mixture of decalin and cyclohexyl bromide (``dec-CHB'') \cite{yethiraj_colloidal_2003}, the Peclet numbers are low, but the interactions can at best be described as ``hard-sphere-like'' with an effective hard-sphere diameter that is $\sim$ 10 to 15\% larger than the measured diameter. The present system is thus a useful find, as such a system can make closer contact with computer simulations; we thus recommend this system as an appropriate model system for precise comparisons of experiments and simulations of hard-sphere-plus-dipolar colloids.

Finally, an interesting aspect of the cluster state might have relevance to the
low-density networking forming ``void'' phase, which are seen in experiments \cite{kumar_new_2005, agarwal_low-density_2009} but not in simulations \cite{almudallal_simulation_2011, richardi_low_2011, park_electric-field-induced_2011}. While the intra-chain structure is thermal at dimensionless dipolar strengths $\Lambda \sim 1$, 
the dynamics of chain-chain clusters on large (time and length) scales is likely to be non-Brownian. Simulations of dipolar colloids in the presence of gravity would shed light on this matter.

\section*{Methods}

\label{sec:sed2eos}

The system used is a colloidal suspension of PMMA colloids in a cis-trans-decahydronaphthalene (decalin) and tetrachloroethylene (TCE) solvent. 
Two particle diameters were used in the experiments. For both particle diameters ($2a = 1.0$ and $0.8 \mu$m), the solvent mixture was adjusted to volume ratios of 60:40 and 70:30 of decalin to TCE respectively (Table~\ref{tab:data}) so that the density mismatch yielded a gravitational length much greater than the particle diameter, $l_{g} \sim 11 \mu$m. 
Density match of the PMMA microspheres and the medium could be achieved with a 50:50 cis-trans decahydronapthalin (decalin) / tetrachloroethylene (TCE) solvent mixture; 
in this case the gravitational length, $l_{g}$ was significantly larger than the sample thickness. 
The $2a = 1.0$ and $0.8 \mu$m particles are dyed with 4-methylaminoethylmethacrylate-7-nitrobenzo-2-oxa-1,3-diazol (NBD) and 1,1-dioctadecyl-3,3,3,3-tetramethylindocarbocyanine  perchlorate  (DiIC18) respectively, and excited by 490nm (blue) and 561 nm (green) laser lines. In the binary mixture, while the emission from the green excitation comes purely from the $0.8 \mu$m particles, the emission from the blue excitation also has a small contribution from the 1.0 and $0.8 \mu$m particles.  
Shown in Table~\ref{tab:data} are the relevant parameters for confocal microscopy, density and refractive index, as well as the relevant parameter for electric field studies, the dielectric constant. The dielectric constant is roughly constant between DC and 1 MHz (AC) but very different at optical frequencies, hence it is feasible to have optical match but electrical mismatch between particles and solvent mixture. It should be noted that van der Waals attractions can be reduced, but by no means made to disappear, simply by matching refractive index.

For all the electric-field cells, the top and bottom plates were 22 mm square cover slips coated with a transparent conducting layer  of indium tin oxide (ITO). These cover slides were glued together (with Norland optical epoxy NOA 68) with two sets of 50 $\mu$m polyethylene terephthalate films placed in between as spacers to give a sample thickness of 100 $\mu$m. A channel inside the cell was filled with colloidal suspension and sealed with the UV glue.
\begin{figure}
     \centering
	\includegraphics[width=160mm]{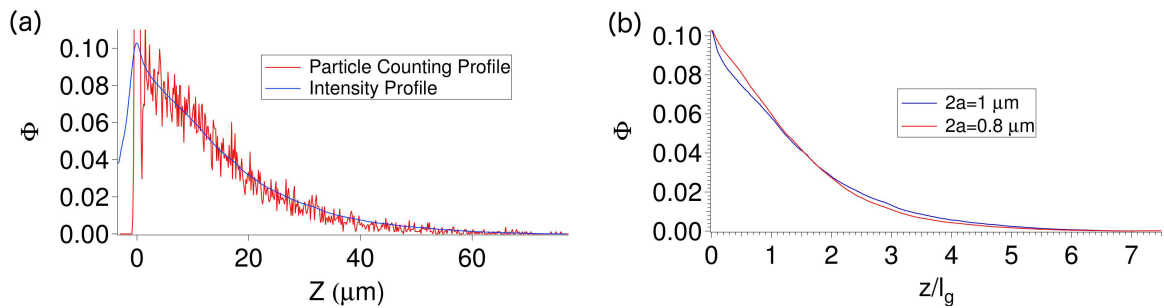}
    \caption{(a) Zero-field sedimentation profiles obtained from particle counting and summed intensity for $1\mu m$ diameter PMMA colloids in 60:40 decalin/TCE compared.
(b) Sedimentation profiles $\Phi$ plotted against the dimensionless height $z/\ell_g$ for two particle diameters ($2a = 0.8$ and $1.0 \mu$m) are essentially identical.}
\label{fig:profiles}
\end{figure}
The colloidal sediment is imaged by rapidly acquiring a z-stack ($77 \mu m$ x $77 \mu m$ x $108 \mu m$) at 29 frames per second ($fps$) using a Visitech VT Eye confocal scanner. Slices in z are acquired in 0.15$\mu m$ z-increments. Standard particle-tracking and counting methods \cite{crocker_methods_1996} can then be used to get the particle density; however we find that using the summed intensity of each image in the z-stack~\cite{beckham_interfacial_2007} shows good agreement with the particle counting technique, as seen in Figure~\ref{fig:profiles}(a). The intensity-based sedimentation profile (Figure~\ref{fig:profiles}(b)), being smoother, is then used to obtain the equation of state (EOS). Clusters are identified simply $via$ proximity, with two particles that are within $1.1 \times 2a$ being identified as connected.

Time series are obtained at several depths $z$ within the sample at capture rates of 29 $fps$ or 56 $fps$. From these time series mean-squared displacements versus time are obtained, and fitted to obtain diffusion coefficients. The time of seconds, in colloids with reasonably high local densities, is in the long-time limit.

\section*{Acknowledgments}
This work was supported by the Natural Sciences and Engineering Research Council of Canada. We acknowledge fruitful discussions with Francesco Piazza, Eli Sloutskin and Ivan Saika-Voivod; we also thank F.P. for critiquing the manuscript. 

\section*{Author information}
\subsection*{Contributions}
H.N. performed all the experiments. H.N. and A.Y. analyzed the results. A.Y. wrote the manuscript. Both authors reviewed the manuscript.
\subsection*{Competing financial interests}
The authors declare no competing financial interests.
\subsection*{Corresponding author}
Correspondence to: Anand Yethiraj

\newpage
\begin{table}
\centering
\begin{tabular}{ | l | l | l | l | l |}
\hline
  & PMMA & 50:50 & 60:40 & 70:30 \\ \hline
$\rho$ ($\mathrm{kg/m^3}$) & 1259 & 1259 & 1186.4 & 1113.8 \\ \hline
$n$  & 1.4895 & 1.4895 & 1.4864 & 1.4833 \\ \hline
$\epsilon_{r}$ & 2.6 & 2.35 & 2.32 & 2.29 \\ \hline
$l_{g}$ ($\mu$m) & - & $\gg 100$ & 10.9 & 10.7  \\ \hline
$2a$ ($\mu$m) & - & $ 1.0$ & 1.0 & 0.8  \\ \hline
$Pe = a/l_{g}$ & - & $\ll 0.005$ & 0.046 & 0.037  \\ \hline 
\end{tabular}
\caption{Density, index of refraction, and dielectric constants for the PMMA particles and three solvent mixtures composed of 50:50. 60:40, and 70:30 (by volume) ratios of decalin and TCE. In addition, the calculated gravitational lengths used in the corresponding experiments are listed.}
\label{tab:data}
\end{table}
We examined the system's response in an applied AC electric field, at a frequency of 1MHz where 
a dielectric response is expected, 
which may be considered to be instantaneous on our slow microscopy time scales. In past work, this response has been modelled as a point dipole at the centre of the sphere~\cite{winslow_induced_1949, tao_electric_1989, tao_electric_1991, martin_structure_1998}, with the dipolar strength depending on the applied electric field according to Equation \ref{eqn:dipstr}; values are shown for different particle sizes in Table \ref{tab:lambda}.
\begin{table}
\centering
\begin{tabular}{ | l | l || l | l |}
\hline
$2a=0.8\mu m$ & & $2a=1\mu m$ &  \\ \hline
E (V/mm)  & $\Lambda$ & E (V/mm) & $\Lambda$ \\ \hline
166.7 & 0.02626 & 92.6 & 0.01270\\ \hline
333.3 & 0.1050  & 185.2 & 0.05080 \\ \hline
500.0 & 0.2363  & 277.8 & 0.1143  \\ \hline
666.7 & 0.4201  & 370.4 & 0.2032  \\ \hline
833.3 & 0.6563  & 463.0 & 0.3175 \\ \hline
1000.0 & 0.9452  & 555.6 & 0.4572  \\ \hline
1166.7 & 1.287  & 648.1 & 0.6221  \\ \hline
1333.3 & 1.680  & 740.7 & 0.8126  \\ \hline
1500.0 & 2.127  & 833.3 & 1.028  \\ \hline
1666.7 & 2.626  & 926.0 & 1.270  \\ \hline
- & - & 1018.5 & 1.53646 \\ \hline
- & -  & 1111.1 & 1.829  \\ \hline
- & -  & 1203.7 & 2.146\\ \hline
\end{tabular}
\caption{Dipolar strength parameter $\Lambda$ for each experiment calculated from the corresponding electric field strength. In order to calculate $\Lambda$, we use dielectric constants from Table~\ref{tab:data}, and equation~\ref{eqn:dipstr}. The 1 $\mu$m-diameter and 0.8 $\mu$m-particles are suspended in 60:40 and 70:30 decalin:TCE respectively in order to have comparable gravitational lengths, thus yielding comparable extents in the sediment.}
\label{tab:lambda}
\end{table}

%

\end{document}